%% file: main.tex
\documentclass[fleqn,usenatbib]{mnras}
\pdfoutput=1  

\usepackage{graphicx}            
\usepackage{amsmath}             
\usepackage{amssymb}             
\usepackage{gensymb}             
\usepackage{bm}                  
\usepackage{xspace}              
\usepackage{nicefrac}            
\usepackage{hyperref}            
\usepackage{xcolor}              
\usepackage{ifthen}
\usepackage{enumitem}
\usepackage{subfig}
\usepackage{suffix}
\usepackage{textcomp}
\usepackage{fontawesome}         

\title[A fossil group at redshift 0.6?]
      {The lens SW05 J143454.4+522850: a fossil group at redshift 0.6?}

\newcommand{\nauthor}[2]{#2,$^{#1}$}
\newcommand{\nauthorl}[2]{#2$^{#1}$}
\newcommand{\email}[1]{\thanks{Email: #1}}
\newcommand{\naffiliation}[2]{$^{#1}$#2}

\newcommand{\SW}{Space\,Warps\xspace}
\newcommand{\Gyrs}{\ensuremath{\mathrm{Gyr}}\xspace}

\newcommand{\Ho}{\ensuremath{H_0}}

\def\pwidth{.46\textwidth}
\def\qwidth{.24\textwidth}
\def\qqwidth{.275\textwidth}

\newcommand{\seclbl}[1]{\label{sec:#1}}

\newcommand{\figlbl}[1]{\label{fig:#1}}

\newcommand{\eqlbl}[1]{\label{eq:#1}}

\newcommand{\secref}[1]{Section~\ref{sec:#1}}
\WithSuffix\newcommand\secref*[1]{Section~\ref{sec:#1}}
\newcommand{\subsecref}[1]{Subsection~\ref{subsec:#1}}
\WithSuffix\newcommand\subsecref*[1]{\ref{subsec:#1}}
\newcommand{\figref}[1]{Fig.~\ref{fig:#1}}
\WithSuffix\newcommand\figref*[1]{\ref{fig:#1}}
\newcommand{\tabref}[1]{Table~\ref{tab:#1}}
\WithSuffix\newcommand\tabref*[1]{\ref{tab:#1}}
\renewcommand{\eqref}[1]{Eq.~(\ref{eq:#1})}
\WithSuffix\newcommand\eqref*[1]{(\ref{eq:#1})}

\newcommand{\Code}[1]{\texttt{#1}}

\newcommand*{\home}{./}%

\author[Denzel et al.]{%
  \nauthor{1,2}{Philipp Denzel}    \email{phdenzel@physik.uzh.ch}
  \nauthor{2}{Onur \c{C}atmabacak}
  \nauthor{3}{Jonathan Coles}
  \nauthor{4}{Claude Cornen}  \newauthor 
  \nauthor{2}{Robert Feldmann}
  \nauthor{5,6,7}{Ignacio Ferreras}
  \nauthor{8}{Xanthe Gwyn Palmer} \newauthor
  \nauthor{1}{Rafael K\"ung}  
  \nauthor{9}{Dominik Leier}
  \nauthor{1,2}{Prasenjit Saha}   
  \nauthorl{8}{Aprajita Verma}
  \newauthor
  \\
  \naffiliation{1}{Physik-Institut, University of Zurich, 8057 Zurich, Switzerland} \\
  \naffiliation{2}{Institute for Computational Science, University of Zurich,
    8057 Zurich, Switzerland} \\
  \naffiliation{3}{Physik-Department, Technische Universit\"at M\"unchen,
    Ernst-Otto-Fischer-Str. 8, 85748 Garching, Germany} \\
  \naffiliation{4}{Zooniverse, c/o Astrophysics Department,
    University of Oxford, Oxford OX1~3RH, UK} \\
  \naffiliation{5}{Instituto de Astrof{\'i}sica de Canarias, Calle V{\'i}a L{\'a}ctea s/n, E38205, La Laguna, Tenerife, Spain}\\
  \naffiliation{6}{Department of Physics and Astronomy, University College London, London WC1E 6BT, UK}\\
  \naffiliation{7}{Departamento de Astrof\'\i sica, Universidad de La Laguna, E38206 La Laguna, Tenerife, Spain}\\   
  \naffiliation{8}{Sub-department of Astrophysics, University of Oxford,
    Denys Wilkinson Building, Keble Road, Oxford, OX1 3RH, UK} \\
  \naffiliation{9}{Dipartimento di Fisica e Astronomia, Alma Mater Studiorum Universit\`{a} di Bologna,
  Viale B. Pichat 6/2, 40127, Bologna, Italy}
}

\date{}
\pubyear{2021}

\begin{document}
\label{firstpage}
\pagerange{\pageref{firstpage}--\pageref{lastpage}}

\maketitle

\begin{abstract}
Fossil groups are considered the end product of natural galaxy group
evolution in which group members sink towards the centre of the
gravitational potential due to dynamical friction, merging into a
single, massive, and X-ray bright elliptical.  Since gravitational
lensing depends on the mass of a foreground object, its mass
concentration, and distance to the observer, we can expect lensing
effects of such fossil groups to be particularly strong. This paper
explores the exceptional system J143454.4$+$522850.  We combine
gravitational lensing with stellar population-synthesis to separate
the total mass of the lens into stars and dark matter.  The enclosed
mass profiles are contrasted with state-of-the-art galaxy formation
simulations, to conclude that SW05 is likely a fossil group with a
high stellar to dark matter mass fraction (0.027$\pm$0.003) with
respect to expectations from abundance matching (0.012$\pm$0.004),
indicative of a more efficient conversion of gas into stars in fossil groups.
\end{abstract}


\input{tex/intro.tex}

\input{tex/system.tex}

\input{tex/lensmass.tex}

\input{tex/stelmass.tex}

\input{tex/simul.tex}

\input{tex/disqus.tex}

%
\section*{Acknowledgments}
PhD acknowledges support from the Swiss National Science Foundation.

This work is based on observations obtained with MegaPrime/MegaCam, a joint project of CFHT and CEA/IRFU, at the Canada-France-Hawaii Telescope (CFHT) which is operated by the National Research Council (NRC) of Canada, the Institut National des Sciences de l'Univers of the Centre National de la Recherche Scientifique (CNRS) of France, and the University of Hawaii. This research used the facilities of the Canadian Astronomy Data Centre operated by the National Research Council of Canada with the support of the Canadian Space Agency. CFHTLenS data processing was made possible thanks to significant computing support from the NSERC Research Tools and Instruments grant program.

\section*{Data availability}
The data underlying this article are available at the Canadian Astronomy Data Centre (https://www.cadc-ccda.hia-iha.nrc-cnrc.gc.ca/; details are cited in the acknowledgements).
The derived data generated in this research will be shared on request to the corresponding author.

%
\bibliographystyle{mnras}
\bibliography{swrefs}

%
\label{lastpage}

\input{tex/appendix}

\end{document}

%% file: tex/intro.tex
%
\section{Introduction}\seclbl{fossil:intro}

The currently most tested cosmological concordance model
($\Lambda$CDM) provides initial conditions for the formation and
evolution of galaxies.  Large-scale cosmological simulations serve as
a framework which explores different galaxy formation scenarios within
$\Lambda$CDM from first principles.  Although these models differ in
their treatment of the baryonic components and the physical mechanisms
involved in galaxy formation, the latest generation of simulation
suites such as EAGLE \citep[Evolution and Assembly of GaLaxies and
  their Environments;][]{Crain15}, FIRE \citep[Feedback In Realistic
  Environments;][]{FIRE}, and Illustris \citep{Illustris} make
remarkably accurate predictions.  In particular, they stand in general
agreement regarding the growth of primordial density fluctuations by
gravitational instability in an expanding Universe, leading to the
formation of dark halos.

In direct relation to the interplay between the structure of the dark
matter distribution and the baryon physics, galaxies are found in a
wide range of structural hierarchies, from low density regions to
groups and clusters \citep[see, e.g.,][]{Tully87,Berlind:06,Yang:07},
and during their lifetime they experience merging events
\citep[e.g.][]{Mergers88, Mergers17}.  In some cases, the mergers
eventually devoid their entire neighbourhood, leaving behind a single
elliptical galaxy of group-scale mass, called a fossil group galaxy
\citep{Fossil94, FossilCenter}. Numerical simulations by
\citet{Barnes89} first motivated such a hierarchical merging scenario
\citep[see also][]{FossilMillenium}.  Since then, there have been
several supporting reports of X-ray sources identified as fossil
groups \citep{FossilSurvey, FossilProps}.  Because most fossil systems
found to date lie within $z < 0.2$, fossil galaxy groups most likely
are old, undisturbed systems due to the lack of major mergers.  While
some luminous galaxies experience major merger events in their
evolution, fossil group galaxies acquire their mass typically through
minor merger events, where the mass ratio stays below 0.3
\citep{Dong:05}.  Simulations show that a fossil system may assemble
half of its mass in dark matter by redshift $z > 1$, and that the
assembled mass at any redshift is generally higher in a fossil than in
regular groups \citep{Dariush07}.  Since this merging process is
relatively fast compared to the cooling time of the surrounding gas,
comparable to one to several Hubble times, fossil groups are usually
found embedded in giant, X-ray luminous gas halos \citep{XRayProps}.
If a fossil system has not yet fully merged, it can be identified by another criterion, a gap in brightness of at least 2
magnitudes (usually defined in the \textit{r}-band) between the two
brightest galaxies in the group \citep{FossilMagGap,Zarattini14}.

As fossil groups are very massive with a high mass concentration, they
can produce strong lensing signatures of background sources
\citep{LensingDiscovery}.  For instance, a massive central group galaxy
with mass $\mathrm{M}\sim10^{13}\,\mathrm{M}_{\odot}$ at a
cosmological distance of $D\sim1\,\mathrm{Gpc}$ features an Einstein
radius of $(4G\mathrm{M}/(c^2D))^{1/2}\simeq10\arcsec$, which is
relatively large compared to typical image separations in single
galaxy lenses, which are of order of sub arc-seconds.  Strong
gravitational lens systems
make for exciting, but rare tools for astronomers to independently
study galaxies.  The morphology of lens systems and their image
configurations allow one to infer mass contents and surface density
profiles \citep[see, e.g.,][]{TK:04,FSW:05,SLACS:08,Auger10,Leier11,
  Whitaker14, Leier16, DEScluster17, Nightingale19}.  On that note, it
is important to realize that the gravitational deflection of light is
independent of the nature of the matter, that is, lensing is equally
affected by baryonic and dark matter.  While lens models of various
forms have been used to constrain mass contents, stellar mass
fractions, dark matter profiles, and even cosmological parameters,
so far these estimates have not yet been translated to constraints on
galaxy formation scenarios.
A recent study by \citet{LensingFossil} estimates that a
substantial percentage of lensing systems within the group mass range
are fossil or pre-fossil groups.  If this is indeed the case, models
of such lenses could provide a new and independent diagnostic -- besides
the common criteria to identify fossil groups through X-ray
observations -- to assess whether a lens in question is in fact a 
fossil system, in contrast to a more standard group.  Thus, models of
such lenses could potentially yield constraints on galaxy formation
scenarios.

In this paper, we present models for the system J143454.4$+$522850
(SW05) of the lensing and stellar mass content, followed by a
comparison with simulations, which indicate several differences
compared with regular early-type galaxies, suggesting the system may
indeed be a fossil group.

The paper is organized as follows.  In \secref{fossil:system}, the
lens system is introduced in detail and its environment thoroughly
investigated.  \secref{fossil:lensmass} presents free-form lens models
of SW05 and several derived diagnostics in order to obtain a
goodness-of-fit estimation.  \secref{fossil:stelmass} describes the
methodology yielding light-to-stellar mass estimates from stellar
population synthesis models and consequently a spatially 2D-resolved
stellar-to-lens mass fraction map.  In \secref{fossil:simul} these
models are compared to simulated galaxies of similar mass range and
\secref{fossil:disqus} summarizes the results and discusses the
feasibility of SW05 as a fossil group galaxy candidate.
Throughout this paper, we adopt a flat cosmology with
$(\Omega_m, \Omega_\Lambda, \Ho^{-1}) = (0.28, 0.72, 13.7\,\Gyrs)$.

%% file: tex/system.tex
\section{The system J143454.4+522850}\seclbl{fossil:system}

  J143454.4$+$522850 or SW05 was discovered in the \SW\ citizen-science
project \citep{SW2}.  Out of the 29 promising lens candidates in that
work, SW05 probably has the best lens image quality.  It is a
relatively large gravitational lens with four clearly separated images
within a radial distance between $3.5$ and $5.25\,\mathrm{arcsec}$
from the centre (see \figref{fossil:composite}).  Early models using the
image positions suggested a typical four-image galaxy lens (see
Appendix~A).  Subsequently, \citet{Kueng18} identified SW05 as a
high-mass galaxy with $\sim10^{13}\,\mathrm{M}_{\odot}$ and a
comparatively high stellar-mass fraction with respect to other
galaxies with similar mass.  The SW05 lens thus sits at the highest
end of the stellar mass distribution of galaxies, making it a very
interesting target for follow-up studies.

\input{fig/composite}

\input{fig/nbrhood}

Originally, the redshift of the lensing galaxy and the background
source were determined with the widely used Bayesian photometric
redshift estimator
\Code{bpz}\footnote{\url{http://www.stsci.edu/~dcoe/BPZ/}}
\citep{bpz}. The code fits the spectral energy distribution (SED)
comprising multi-band photometry with a set of well-calibrated
templates. The redshift derived by \Code{bpz} for SW05 is $z_L =
0.63\pm0.16$. We later cross-matched the sample with Data Release 16
of the Sloan Digital Sky survey \citep{SDSSDR16} and found a match for
the lens in the BOSS survey \citep{BOSS} at $z_L=0.62525\pm
0.00020$. The spectrum shows prominent Ca{\sc II} H+K lines along with
a substantial 4,000\AA\ break, representative of an old stellar
population. In contrast, \Code{bpz} was not able to determine the
redshift of the source with enough credibility such that a default
redshift from the MegaPipe pipeline was used for the lens model at
$z_S = 3.00$.  Accurate redshifts are not essential for the lensing
models, as a change in redshift (of the lens and the source) can be
implemented by rescaling the critical surface density, and can
therefore easily be adjusted to any other redshift.  Moreover, the
lens models for SW05 seem to be rather insensitive to changes in the
source redshift.  Contrarily, a change in lens redshift can
considerably impact the results of the photometry-based analysis for
the stellar-mass estimation.  During the analysis of SW05,
spectroscopic redshifts were measured (by XGP and AV) for the source,
at around $z_S=2.6$, and also confirmed the lens redshift.
The spectroscopic data was taken with Oxford's SWIFT spectrograph mounted
at the Hale Telescope at the Palomar Observatory.

\figref{fossil:nbrhood} shows the wider field of SW05. The left panel
shows a $\sim 4\times 4$\,arcmin$^2$ colour image from the CFHTLS data
in the $g$, $r$, and $i$ filters, with SW05 at the centre. At the lens
plane, the image roughly maps the region within 0.8\,Mpc of SW05.  The
right panel shows the sources with a confirmed redshift. The green
dots mark galaxies within $\Delta z=\pm 0.1$ of the redshift of SW05,
i.e. potential group members (with symbol size scaling with flux).
The other dots correspond to sources that cannot be associated to the
group of SW05, being either blue- or redshifted with respect to SW05
(coloured accordingly).  The grey symbols are stars. A circumference
with radius 0.5\,Mpc at the lens plane is shown for reference. It is
the first indication for SW05 being a fossil group galaxy candidate as
its neighbourhood is relatively clear of galaxies at similar
brightness and redshift.

%% file: fig/composite.tex
\begin{figure}
  \includegraphics[width=\pwidth]{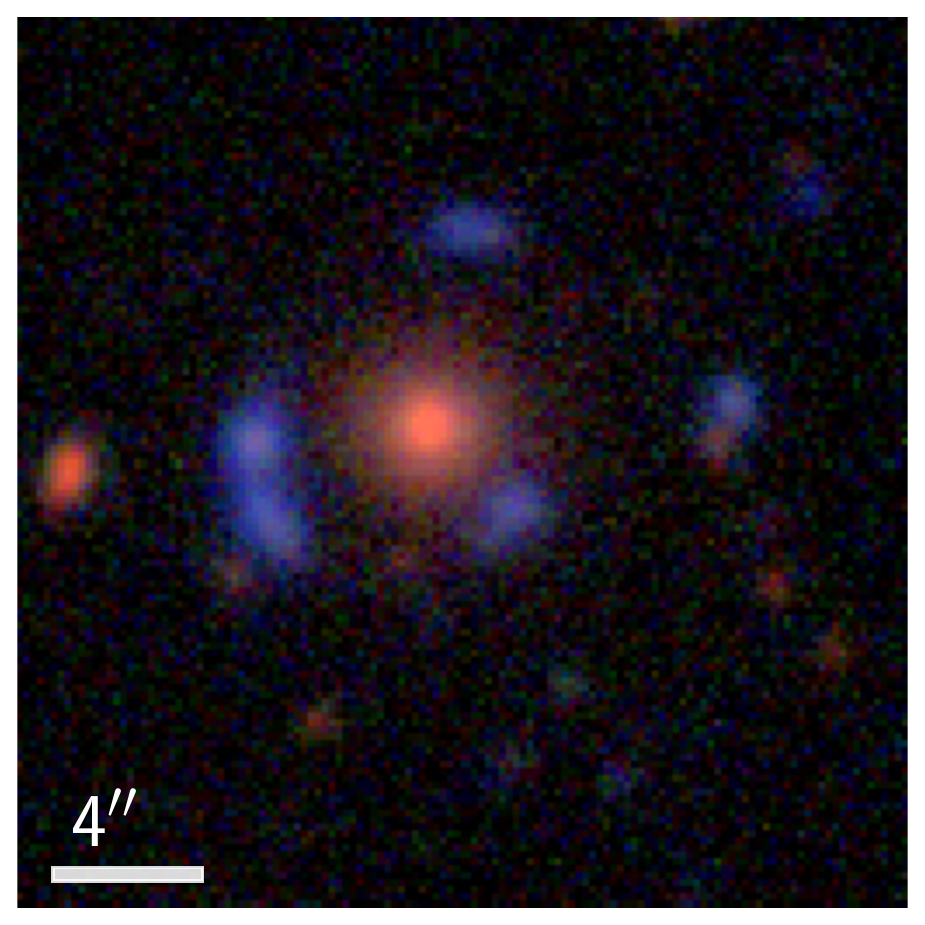}
  \caption[Composite image of SW05]{%
    Stacked observational picture of SW05: Observational data from CFHTLS
    (stored in the CFHT Science Archive) was taken with the wide-field imager
    MegaPrime in five optical bands (u, g, r, i, and z).  The false-colour image
    was generated using a stacking procedure according to
    \protect\cite{Lupton04}, where the i, r, and g bands are transformed into
    rgb colours.}
    \figlbl{fossil:composite}
\end{figure}

%% file: fig/nbrhood.tex
\begin{figure*}
  \includegraphics[width=80mm]{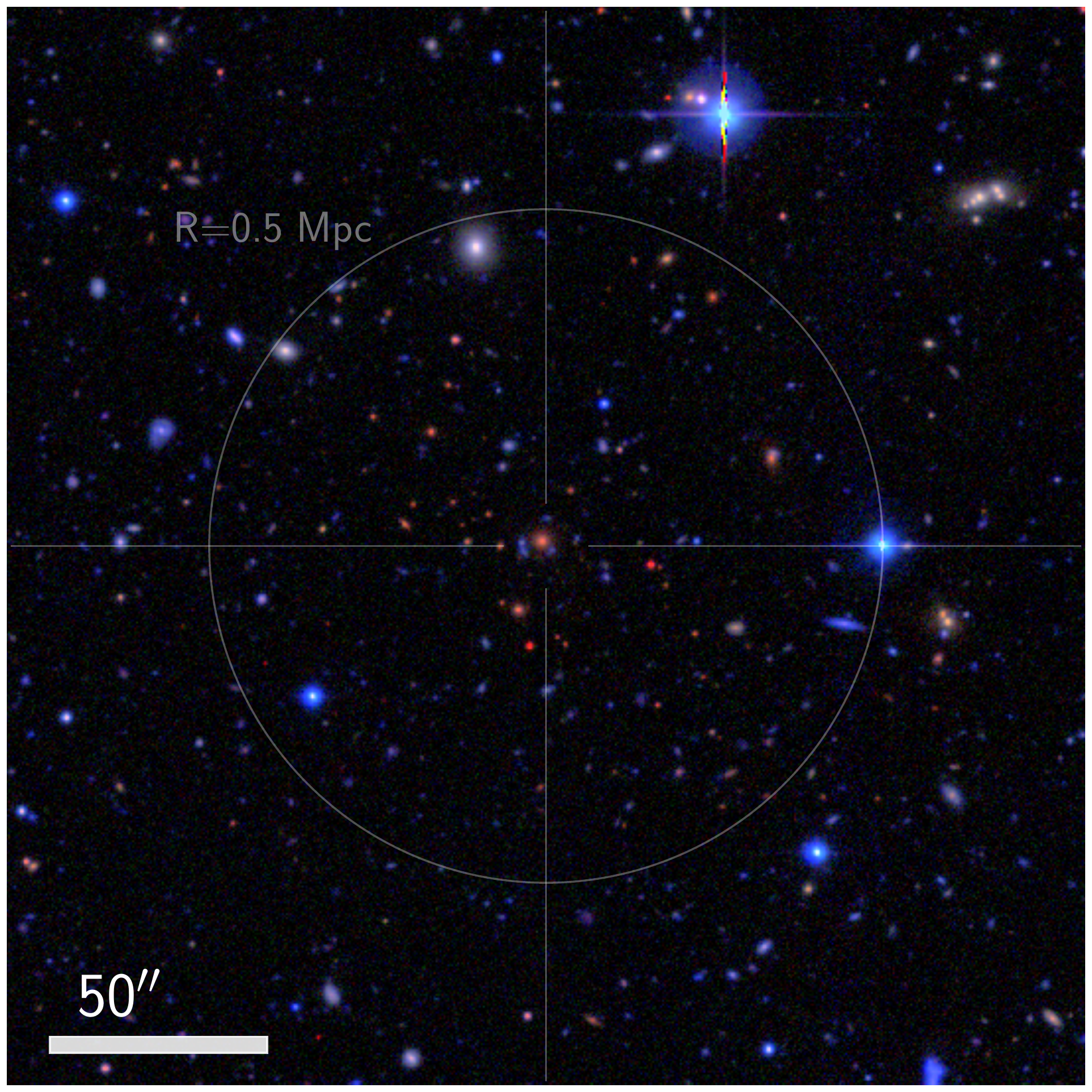}
  \includegraphics[width=80mm]{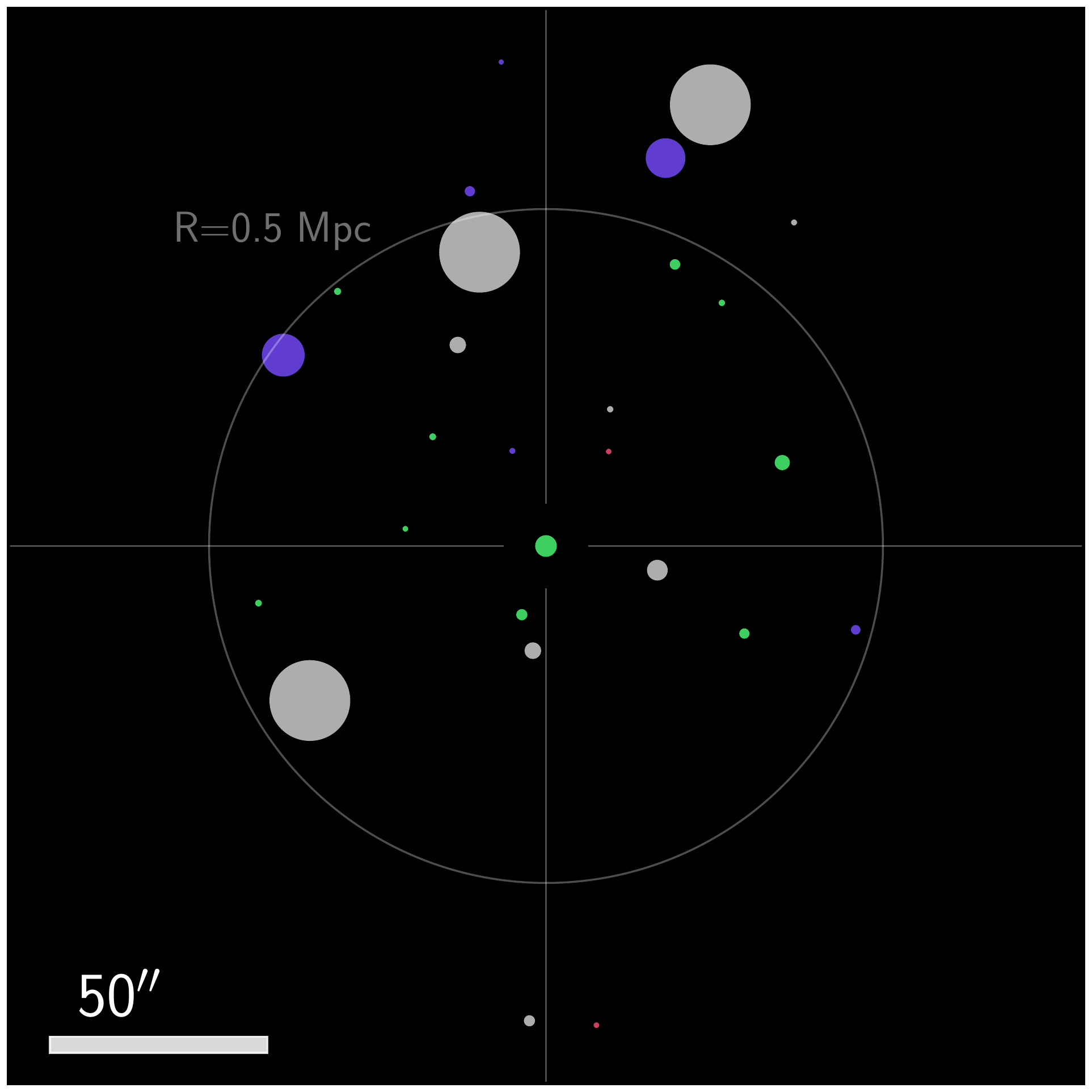}
  \caption[The neighbourhood of SW05]{%
    The left panel shows a composite image of the neighbourhood of SW05 from
    CFHTLS.  All the image parameters, including the stacking procedure is the
    same as for \figref{fossil:composite}.  The panel on the right marks 
    potential group members based on redshift. Green dots represent
    objects within a redshift range of $\pm0.1$ of SW05 (including the lens
    at the centre). Red dots stand for objects further redshifted,
    blue dots for further
    blue-shifted objects.  Grey dots are stars or unspecified sources.
    The size of the dots
    mark the relative brightness.  The scale bar has a length of $50\;\arcsec$
    which corresponds to a linear distance of roughly 330\,kpc at the
    lens redshift (z$_L$=0.625). For reference, a circumference with
    radius 0.5\,Mpc at the lens plane is shown.}\figlbl{fossil:nbrhood}
  \end{figure*}

%% file: tex/lensmass.tex
\section{The lensing mass}\seclbl{fossil:lensmass}

The lens modelling was done with the reliable free-form modelling code
\Code{GLASS}\footnote{\url{https://ascl.net/1806.009}} \citep{GLASS}.
The free-form  technique addresses the 
many degeneracies present in lens models \citep{Saha2000, Saha06}.  In
contrast to conventional parametric models, free-form models allow a 
flexible analysis by constructing the mass distributions $\Sigma(\bm\theta)$ from
a high number of base elements.  \Code{GLASS} -- analogously to the free-form
modelling tool \Code{PixeLens} \citep{PixeLens} -- uses ``mass tiles''
$\Sigma_n$ to construct its lens models according to:
\begin{equation}
  \Sigma(\bm\theta) = \sum_{n} \Sigma_n Q(\bm\theta - \bm\theta_n),
\end{equation}
where $Q$ is a square-pixel profile and $\theta_n$ its
centroid.  The image point positions $\bm\theta$ provide the
linear constraints on the intrinsic source position $\bm\beta$, along with the mass
components $\kappa_n$. Here, the lens is mapped into 14 rings of pixels around its
centre with a total of 497 pixels.  However, this does not uniquely define the
lens mass distribution and in order to obtain reasonable models, their solution
space is further constrained with inequality priors as follows:
\begin{enumerate}
  \item All mass tiles must have non-negative densities: $\Sigma_{n} \geq 0$.
  \item To ensure smoothness in the mass distribution, each mass pixel is
  limited to twice the average of its neighbours.
  \item The local density gradient should point within $\alpha = 45^{\circ}$ of
    radially inwards which assures concentration of the lens and more
    importantly suppresses additional lens images which might appear otherwise.
  \item The average density $\langle\Sigma\rangle_{i}$ of mass within a
    concentric pixel rings is required to not increase with radius.  This still
    allows for twisting iso-density contours and significantly varying
    ellipticities with radius.
\end{enumerate}%

\input{fig/arrival}

Additionally, to account for any mass component outside the finite
model surface, the code allows for a two-component external shear
$\bm\gamma$.  Solutions are then sampled with a customized Monte-Carlo
random-walk method \citep{Lubini12}.  The final model consists of an
ensemble of 1000 solution sets of ($\Sigma_n$, $\bm\beta$,
$\bm\gamma$) for the image point positions.
To verify the validity of the lens models, two main diagnostics are
used.  Firstly, inspection of the arrival-time surface (see
\citealt{Kueng15, Kueng18}) is an ideal test to reject certain models
in the ensemble, as additional images and uneven contours, indications
of unphysical solutions, are easily discernible.  Minima, and
saddle-points, correspond to points where the gradient of the
arrival-time surface is zero and indicate detectable source-image
positions.  Maxima are usually not detectable as they are highly
demagnified.  In general the arrival-time surface should:
\begin{enumerate}%
  \item reproduce the source image positions on the lens plane,
  \item not produce more extrema than the observation shows,
  \item have reasonable delay times relative to each other.
\end{enumerate}%
\figref{fossil:arrival} shows the arrival-time surface of the ensemble-average
model of SW05.

The high sensitivity of the lens models to the exact image positions
is still a troubling issue.  An automated search algorithm is quite
unreliable due to the generally high signal-to-noise ratios in the
data from CFHTLS.  Especially with spatially extended images, such as
arcs, the correct positions are at times arbitrary even in a visual
inspection, and a slight shift can cause the mass model of the lens to
change moderately, at times even drastically.

\input{fig/reconsrc}

Thus, the second diagnostic takes the entire photometric data, rather
than single image points, into account by reconstructing the lensed
source from the image plane using the mass distribution of the
ensemble average solution.  This is done with
\Code{gleam}\footnote{\faGithub\,
  \url{https://github.com/phdenzel/gleam}} (Gravitational Lens
Extended Analysis Module) (see \citealt{Denzel20} for details).
\Code{gleam} uses a simple least-square fitting method to produce
synthetic lens images as shown in \figref{fossil:reconsrc}.  Here, the
photometric data in the $g$, $r$, and $i$ bands were fitted with
synthetic images, with an average reduced $\chi^{2}_{\nu} \sim 2.10$,
and produced synthetic composite images (upper right panel, in
contrast with the actual image, shown upper-left).  The pixel-wise
residuals between observed and synthetic data, scaled by noise, is
shown in the bottom-right panel.  The reconstructed source composite
is shown in the bottom-left panel.  These fits provide excellent
diagnostics and confirm the validity of the underlying mass model, as
the synthetic image is well reproduced in three different bands, and
the reconstructed source is relatively consistent.  While a match of
synthetic images with the observations does not necessarily mean that
the true model from the degenerate solution space has been found, the
inverse argument can still be used to invalidate a solution.

%% file: fig/arrival.tex
\begin{figure}
  \centering
  \includegraphics[width=\pwidth]{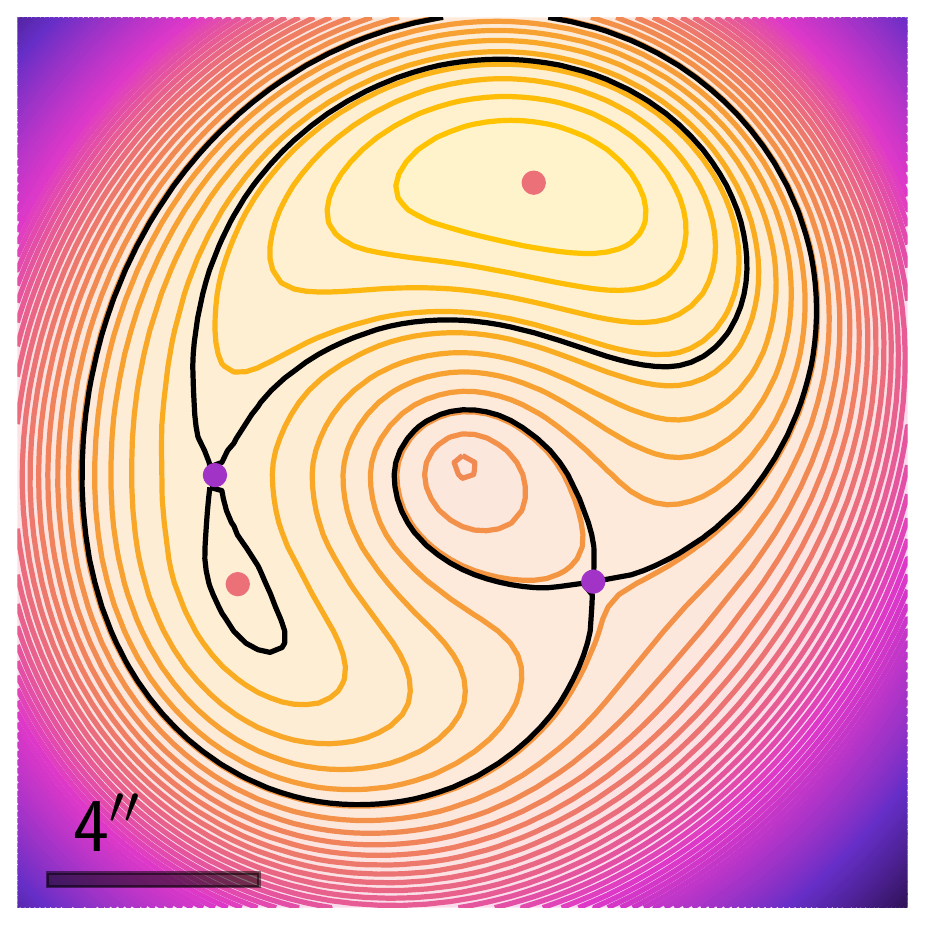}
  \caption[Arrival-time surface of the ensemble-average model for SW05]{%
    Arrival surface of SW05's models.  This figure shows the light travel times for
    virtual light paths from the source to the observer.  Its maxima, minima,
    and saddle-points are an illustration of Fermat's principle, which states
    that arrival times must be extreme for paths along which light rays travel.
    The black lines show the saddle-point contours of the surface, the coloured
    lines general surface contours.  }\figlbl{fossil:arrival}
\end{figure}

%% file: fig/reconsrc.tex
\begin{figure}
  \centering
  \begin{tabular}{@{}l@{}l@{}}
    \includegraphics[width=\qwidth]{\home imgs/composite_SW05}&\includegraphics[width=\qwidth]{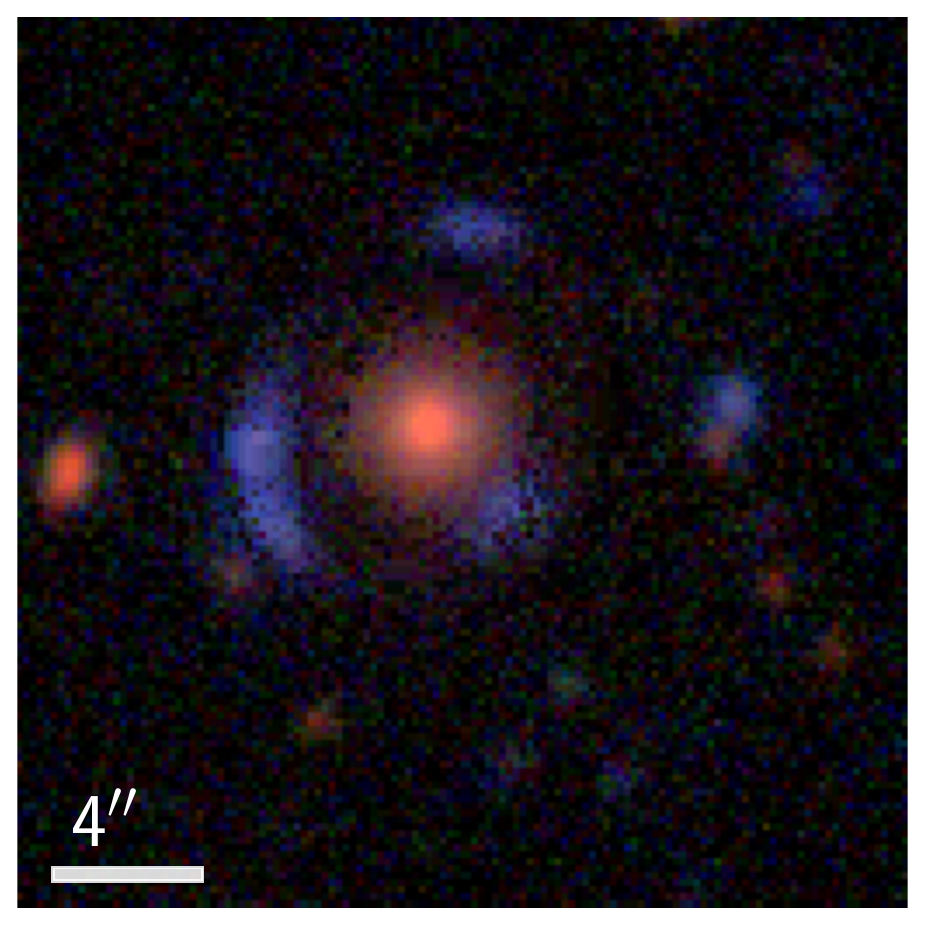} \\
    \includegraphics[width=\qwidth]{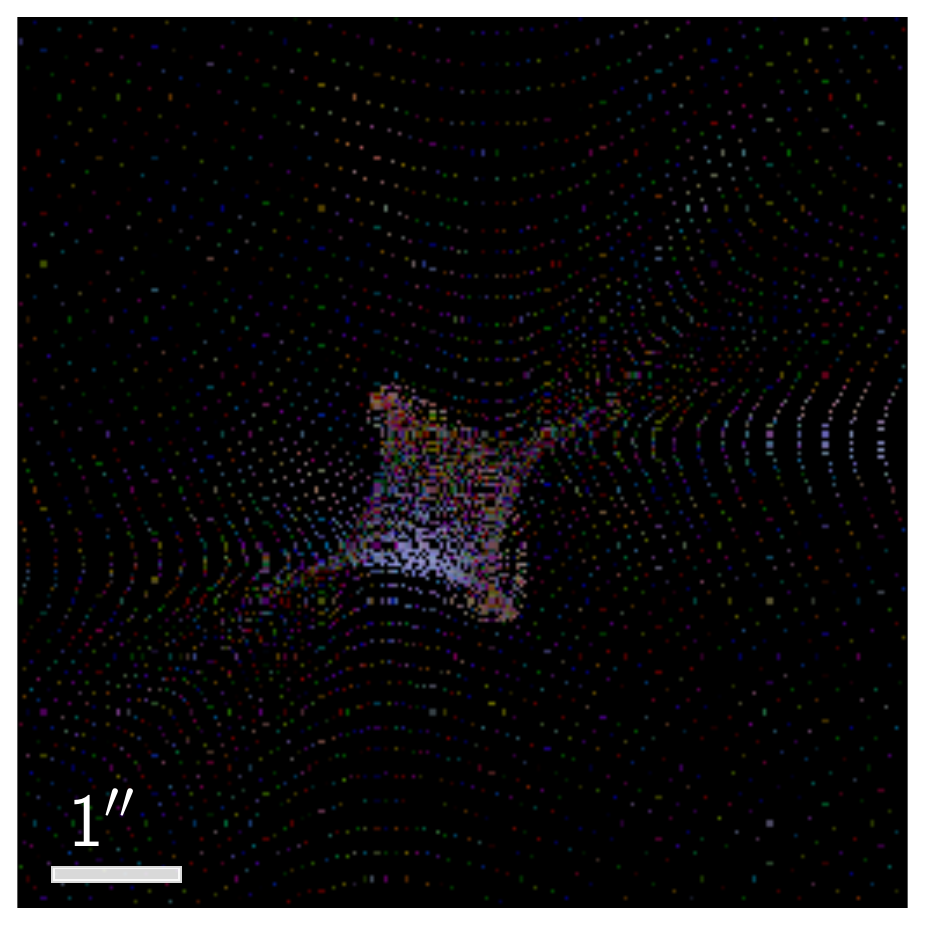}&\includegraphics[width=\qqwidth]{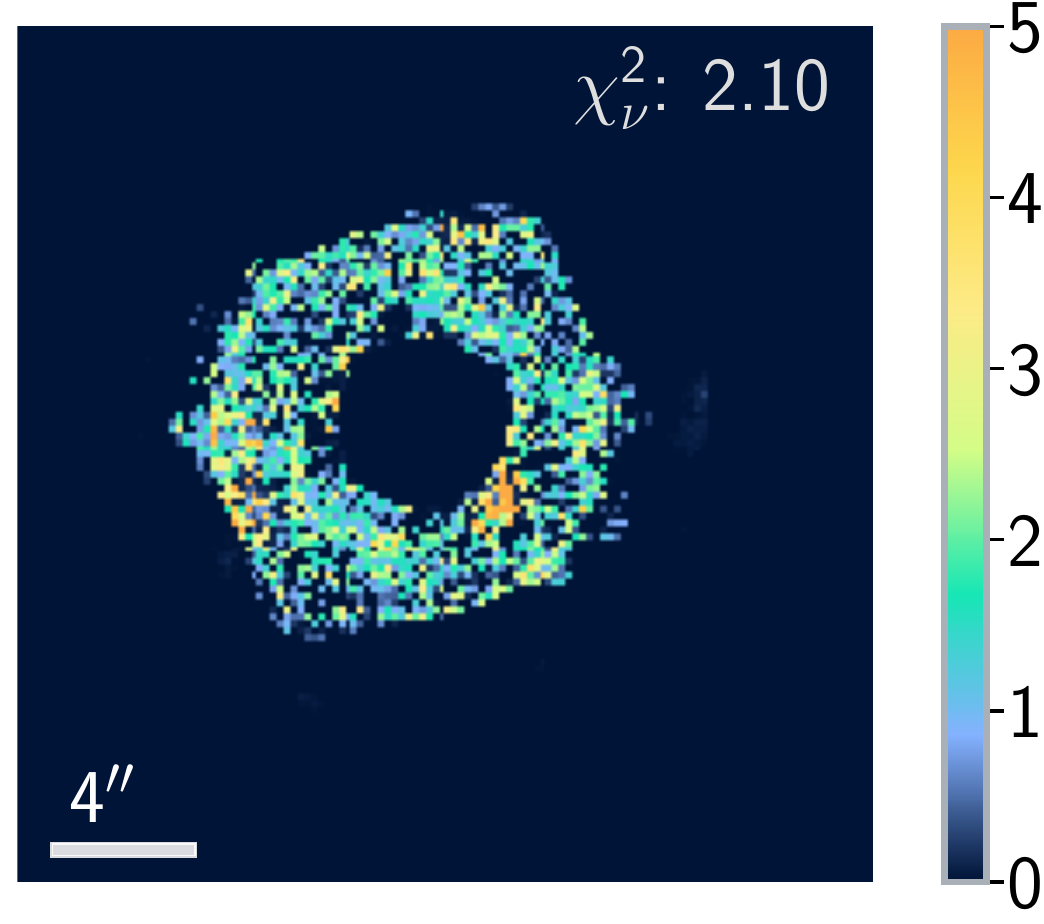}\\
  \end{tabular}
  \caption[Synthetic-images and source fits from the ensemble-average lens
  model]{Synthetic-images and source fits from the ensemble-average lens model.
  Using the modelled lens potential inferred from the mass maps, the observed
  data (top left) is projected onto the source plane to reconstruct the original source (bottom left).  A
  re-projection onto the image plane results in a synthetic image (top right) which provides
  a diagnostic, i.e.\ a visible measure of the models' accurateness when
  compared to the observation.  The bottom-right panel shows pixel-wise
  residuals between observed and synthetic data.}\figlbl{fossil:reconsrc}
\end{figure}

%% file: tex/stelmass.tex
\section{The stellar mass}\seclbl{fossil:stelmass}

The surface brightness models of SW05 were obtained by fitting a
\citet{Sersic} profile to the CFHT Lens images
(\figref{fossil:composite}) via the Markov-chain Monte-Carlo (MCMC)
ensemble sampler \Code{emcee}\footnote{\faGithub\,
  \url{https://github.com/dfm/emcee}} \citep{emcee}.  This feature is
part of the \Code{gleam} module.
In order to translate the photometric data into a stellar mass map,
we obtain the stellar mass-to-light ratio by comparing the colours
with population synthesis models, once more adopting the MCMC sampler
\Code{emcee}. The base models comprise a set of 12 composite stellar
populations from \citet{BruzualCharlot03} that adopt a \citet{ChabrierIMF} 
initial mass function. The 12 models consist of three different choices of
metallicity (assumed constant in each case, [Z/H]={$-0.5, 0, +0.3$}), and
four ``time'' steps, made by assuming a constant star formation rate over
four different time intervals that cover the available time span of SW05.
The fitting code explores the parameter space spanned by all linear
superpositions of the 12 base models, along with an additional dust
attenuation parameter that applies a foreground screen with the Milky Way
standard extinction law \citep{CCM89}. For each realization, the
observed colours from the full CFHT Lens set ${u,g,r,i,z}$ are
compared with the model predictions to produce a likelihood, from
which the best fit and the sampled posterior are used to determine the
mass-to-light ratio.  We note that while the fits to the actual star
formation histories can suffer from degeneracies, the derivation of
M/L is more robust.

\input{fig/lightvsdark}

Finally, the independently derived lensing and stellar mass maps of
SW05 were superimposed with a standard 2D interpolation scheme.  The
result is shown as a stellar-to-lens mass fraction map in
\figref{fossil:lightvsdark}, following a similar colour-coded
presentation of the values and the uncertainties as in \citet{SLACS:08}.
It is a circular 2D false-colour map
with 12 mass tiles --- or pixels --- in radius, which describes the
stellar and total surface densities generated from the ensemble.
By default, the mass models have a range of almost twice the
maximal image separations from the lensing galaxy. Since the image
separation of SW05 is relatively large compared to the other lenses
discovered in the \SW~project, its surface density was mapped with a
larger pixel scale.  For this reason the central region of the lens map
was adaptively refined in order to resolve potential cusps.  The
result is a spatially large map, in which the high concentration of
stars in the centre is clearly evident as shown by the redder region.  The
blue region at larger radii represents lower values of the stellar-to-lens
mass fraction, i.e. where dark-matter is the dominant component.  The
dark matter in the average model mass of SW05 adds  up to
$(1.12\pm0.08)\cdot10^{13}\,\mathrm{M}_{\odot}$, while the stellar
mass amounts to $(3.04\pm0.22)\cdot10^{11}\,\mathrm{M}_{\odot}$.
Therefore, the stellar to dark matter fraction in the SW05 lens
is $0.027 \pm 0.003$. Since the
stellar-to-total mass fraction spans a large range, the mass fraction
was stretched in the false-colour map with a cube root to make subtle
differences in the stellar halo in red colour stretch farther out
(assuming that the stellar mass profile is monotonically decreasing)
and thus easier to see with the human eye.  In general, the galaxy
model shows the expected features.  If there is any high stellar mass
content, it can be found in the centre of the galaxy.  The dark matter
is located in an extended halo that always dominates the total matter
content, especially towards the outskirts of the galaxy, as previously
found in other lensing systems \citep[e.g.][]{FSW:05,Leier11}.

%% file: fig/lightvsdark.tex
\begin{figure}
  \includegraphics[width=\pwidth]{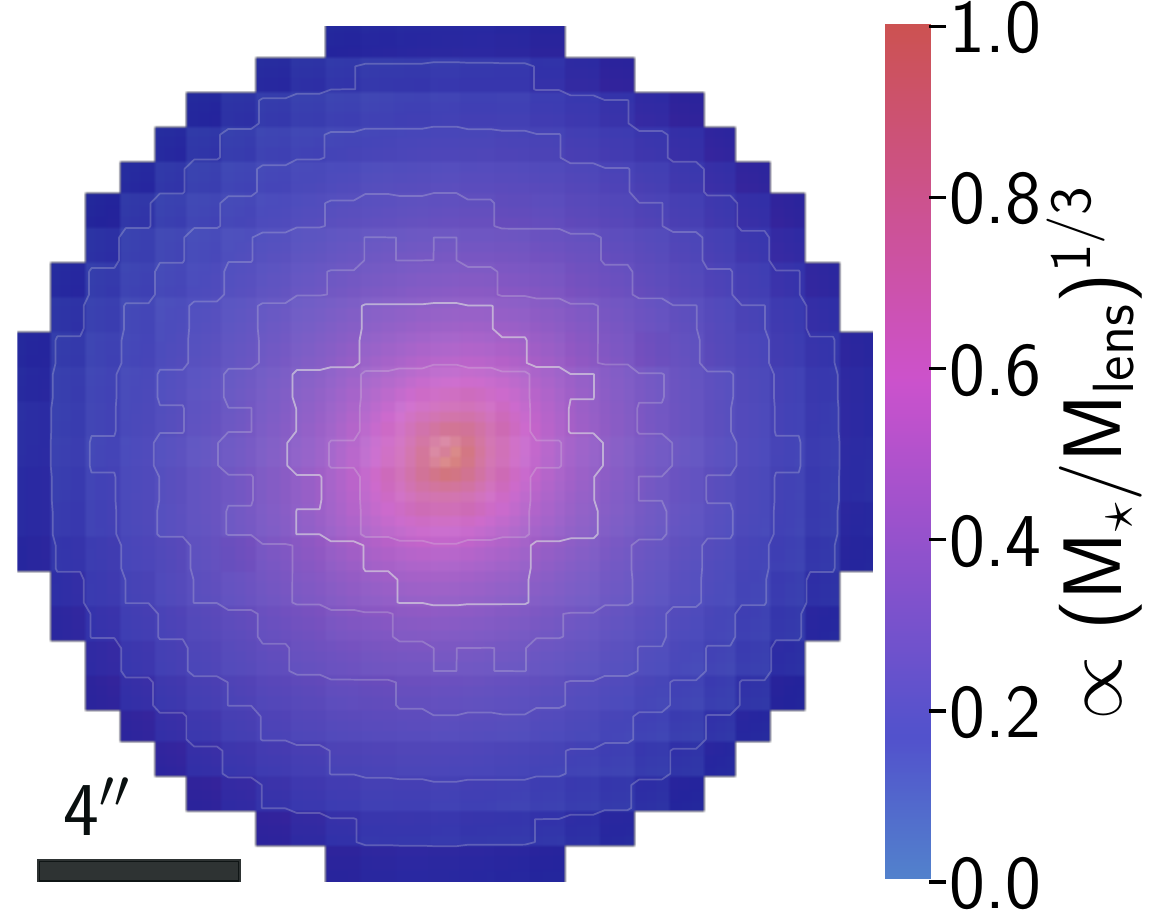}
  \caption[Stellar-to-lens mass surface density map]{False-colour map of
    $f={(\Sigma_{\mathrm{stel}}/\Sigma_{\mathrm{lens}})}^{e_{f}}$,
    i.e.  the fraction of stellar to lensing surface mass density in
    the SW05 lens.  Three components were used for the colour mapping,
    stellar-to-lens surface mass density fraction, namely $f$, the
    lensing surface mass density normalization
    $A={\Sigma_{\mathrm{lens}}}^{e_{A}}$, and the fractional
    uncertainty $\Delta =
    {(\Sigma_{\mathrm{error}}/\Sigma_{\mathrm{lens}})}^{e_{\Delta}}$,
    where $e_{f}$, $e_{A}$, and $e_{\Delta}$ are arbitrarily
    adjustable exponents.  Those are necessary due to the fact that
    the visual perception of colours is highly non-linear.
    With increasing $f$ the colour changes from blue (dominating dark
    matter content) to red (high stellar mass content). With increasing $A$ the
    shading changes from dark (low total mass content) to full colour (high
    total mass content). With increasing $\Delta$, the colour saturation
    changes from full colour (low uncertainty) to white (high uncertainty).  The
    grey lines describe contours of equal surface density, where the brightest
    contour indicates the level of critical surface density, i.e. a convergence
    of 1.  }\figlbl{fossil:lightvsdark}
  \end{figure}

%% file: tex/simul.tex
\section{Comparison with simulations}\seclbl{fossil:simul}

\input{fig/profiles}

We now contrast the mass profile of the SW05 lens with
data corresponding to the formation of massive galaxies in numerical
simulations, to determine whether this is a special system compared
with the typical mass distribution of this type of 
galaxies. \figref{fossil:profiles} shows the cylindrically-averaged
cumulative mass profiles.  The images lie on the lens plane between
22--38\,kpc of the lens centre (vertical dashed lines), and the
Einstein radius sits around the middle of this range.
The projected mass of the lensing galaxy within the Einstein radius is 
$\sim(7.01\pm1.06)\cdot10^{12}\,\mathrm{M}_{\odot}$ in dark matter and
$\sim(2.58\pm0.19)\cdot10^{11}\,\mathrm{M}_{\odot}$ in
stars. At distances of the order of the image separations, the
uncertainties in the lens mass are at their lowest. Away from this
radial position,  the uncertainties grow in both directions, producing
a characteristic butterfly-shaped envelope, caused by the well-known steepness
degeneracy problem \citep{Saha2000, Saha06}. The shape is only
slightly recognizable in this plot, as a logarithmic scaling was used on both
axes.  A factor alleviating the steepness degeneracy is the
comparably high image quality and overall size of the lensing system,
which allows for a precise fix of the positions of the images, which in turn
makes the estimation of the total mass content more accurate.  The
errors in the stellar mass distribution arise from the compatibility
of the observed photometric data at the determined redshift with
multiple base models to different degrees. The choice of IMF also
introduces a systematic uncertainty, but we will see below that
our conclusions are not affected by this.

The simulated galaxies shown in \figref{fossil:profiles} were taken from MassiveFIRE
simulations \citep[A1, A4, C1, D7][]{Feldmann16, Feldmann17}.
These galaxies were selected due to the comparability of their total mass at redshift $0.5<z<0.6$ to SW05.
The simulation profiles were recovered with the halo finder \Code{AMIGA} 
\citep{AMIGA1, AMIGA2}.
The \Code{AMIGA} halo finder provides the output as
spherically averaged profiles of stellar and dark matter particles.
To have a fair comparison however, they were re-averaged within a
cylindrical volume, and re-integrated along the $z$ coordinate to
yield a projected 2D mass profile of the simulated galaxies.  To make
the comparison easier, the simulated profiles were scaled in mass to
have the same total stellar mass, and in radius to have the same
stellar mass at the half-light radius.  Among the simulation profiles
themselves, there are almost no differences except in the total mass.
The enclosed lensing mass profile of SW05 is significantly steeper, and 
the stellar profile shows differences in the central regions, being shallower
than the simulations at R$\sim$1\,kpc. We interpret such differences
as a peculiarity of the lens SW05, being a fossil group, thus representing
a special environment with respect to other galaxies of similar mass.

%% file: fig/profiles.tex
\begin{figure}
  \includegraphics[width=\pwidth]{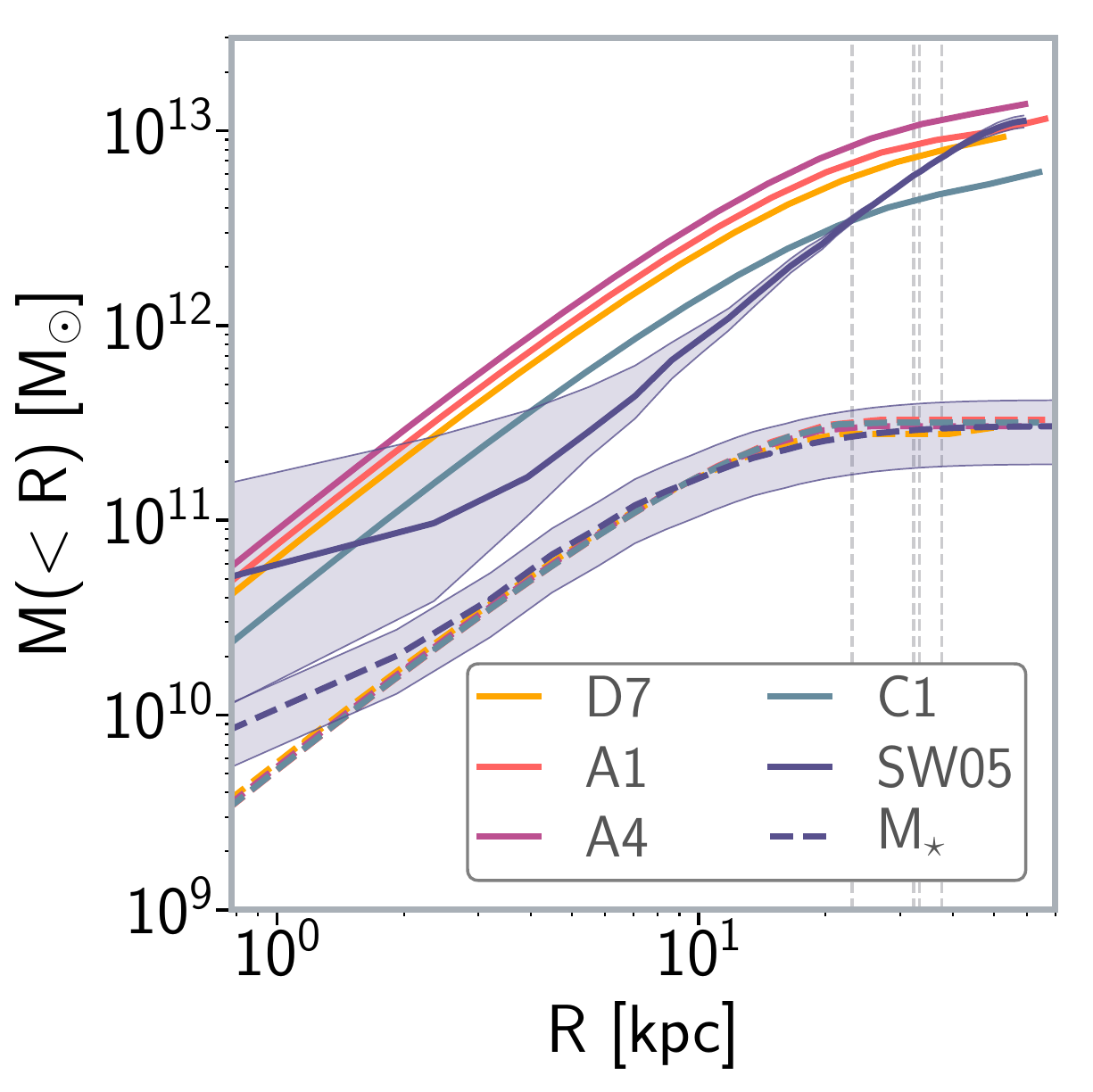}
  \caption[Enclosed mass profiles of SW05 in comparison with FIRE galaxy
    models]{
    Cumulative mass profiles of the SW05 lensing galaxy models
    and galaxies from MassiveFIRE simulations \citep[A1, A4, C1,
      D7][]{Feldmann16, Feldmann17}.  The solid curves (lens model in
    dark-blue) denote the ensemble median of the enclosed total
    mass. The dashed curves (lens model in dark-blue) give the
    ensemble median of the enclosed stellar mass.  The grey areas show
    the 99.7\,\% confidence range of the SW05 lens models.  The
    enclosed lens mass is best constrained at the radial location of
    the images of the background source (vertical thin dotted
    lines).}\figlbl{fossil:profiles}
  \end{figure}

%% file: tex/disqus.tex
\section{Discussion}\seclbl{fossil:disqus}

The results presented in this paper suggest that SW05 is one of the
most massive lensing galaxies known so far, with an Einstein radius
extending to $\sim$5\,arcsec, that maps into $\sim$30\,kpc on the lens
plane. The stellar to dark matter mass budget can be explored in
detail out to the outskirts of the luminous component.  Its mass
estimate fits the expectations of a massive elliptical galaxy, with a
halo mass in the galaxy group range.  Its stellar mass dominates the
centre, and in total it makes up to 2.7\% of the entire mass
budget. These results, along with a lack of nearby bright galaxies
suggest that SW05 is a fossil group.  Fossil groups clear their
neighbourhood from any other group members at early times through
mergers.  If there were any other group members, it would become
evident in the lens models as a external shear component giving rise
to an ellipticity in the mass maps.  In the lens models of SW05 there
is little ellipticity to begin with.  There seems to be a slight
departure from radial symmetry towards the upper left quadrant in
\figref{fossil:lightvsdark}.  However inspecting \figref{fossil:nbrhood}, we
only find one possible group member candidate, which can be found in
the lower left quadrant.  The lens models therefore indicate that
there is no other major group member in the neighbourhood of SW05,
supporting our conclusion.  This explanation also seems to fit with
the discrepancies found in the comparison with numerical simulations
(\figref{fossil:profiles}).

The stellar mass in SW05 accounts for around 2.7\% of the total mass.
However, this is not necessarily equal to the entire baryonic mass in the
lensing galaxy.  There could be a substantial mass content in hot, diffuse
gas, which would radiate in an energy range of
  \begin{align}\eqlbl{fossil:zeldovich}
    \nonumber\left(\frac{GM}{c^{3}}\right)\left(\frac{c}{r}\right) \times 1\,\mathrm{GeV} &\sim
    5\times10^{-8}\,\mathrm{keV}\times\left(\frac{M}{\mathrm{M}_{\odot}}\right)\times
    \left(\frac{r}{\mathrm{pc}}\right)^{-1} \\
    &\sim 10\,\mathrm{keV}.
  \end{align}

This falls in the X-ray spectral window, of which no observational
data was taken so far. Another indication of the special nature of
the SW05 lens can be obtained through a dynamical tracer such as
velocity dispersion. The available SDSS/BOSS spectrum gives a
value for the stellar component, within the 2\,arcsec diameter fiber of the
spectrograph of $\sigma=256\pm 32$\,km/s. Moreover, we can use the
proxy from lensing data proposed by \citet{Leier09}, namely:
\begin{equation}\eqlbl{fossil:dispersion}
   \sigma^{2} = \frac{2}{3\pi} \frac{\mathrm{G}\mathrm{M}(<r)}{r},
\end{equation}
shown in \figref{fossil:dispersion}. 
The figure also shows the values of $\sigma$ for the simulations, that
reach a maximum at smaller galactocentric radii compared to SW05.
This also means that the total mass of SW05 could be well above
$10^{13}\,\mathrm{M}_{\odot}$ if the galaxy model were to be extended to 
larger radii.

\input{fig/dispersion}

The velocity dispersion can formally be converted into a temperature with
\begin{equation}\eqlbl{fossil:gas_temp}
  T(r) = \frac{m_{\mathrm{p}}\sigma(r)^{2}}{e} \quad\text{(eV)}
\end{equation}
where $m_{\mathrm{p}}$ is the proton mass, $\sigma(r)$ the velocity
dispersion profile, and $e$ the elementary charge.  The right $y$-axis
of \figref{fossil:dispersion} shows such a conversion.  This temperature
should be of the same order as the temperature at which the gas in the
outskirts of the galaxy radiates.  The maxima for the observed and
simulated galaxies have values around $\sigma \sim
400\,\,\mathrm{km/s}$ and $T \sim 2\,\,\mathrm{keV}$.

We also contrast the total stellar to dark matter mass ratio found in
SW05 with the standard abundance matching (AM) trends. These trends
are obtained by comparing the halo mass function of numerical
simulations with the stellar mass function from observational
data. Such a comparison allows us to determine the variation in the
stellar to dark matter mass fraction as a function of halo mass, a
signature of overall star formation efficiency. \figref{fossil:AM}
compares our estimate of this fraction for the SW05 lens (1\,$\sigma$
error bars) with the AM relation of \citet{Moster:10}, shown as a line
and a shaded region that encompasses the 1\,$\sigma$ uncertainties of
the trend at the lens redshift (z$_L$=0.625). The AM relation features
a well-known peak at a halo mass around M$_h\sim 10^{12}$M$_\odot$
with a stellar-to-dark matter mass fraction $\sim$2\%, followed by a
decreasing trend with increasing halo mass. At the mass of SW05, the
AM relation gives a stellar to dark matter fraction of 0.012$\pm$0.004, 
well below the observed value for SW05 (0.027$\pm$0.003), suggesting that the
conversion of gas into stars in this galaxy has been significantly
higher than the average, at similar mass, and even higher than the
peak of the relation at the same redshift. Such a result could be
expected within our fossil group hypothesis, if we consider that in
such a special environment -- where merging and/or infall processes lead
to the formation of the central massive galaxy -- took place at very early
cosmic time, when there was larger gas reservoirs to fuel star
formation. This may represent a clear signature to identify fossil
groups.

In conclusion, the study presented in this paper identifies a
rare gravitational lens of group mass-scale as a fossil group candidate
at intermediate redshift, with a significantly high stellar-to dark matter
mass fraction with respect to the expectations from abundance matching.
In the future, this analysis could very well be applied to other
lensing groups to help in the identification of fossil groups.

\input{fig/abmatch}

%% file: fig/dispersion.tex
\begin{figure}
  \includegraphics[width=\pwidth]{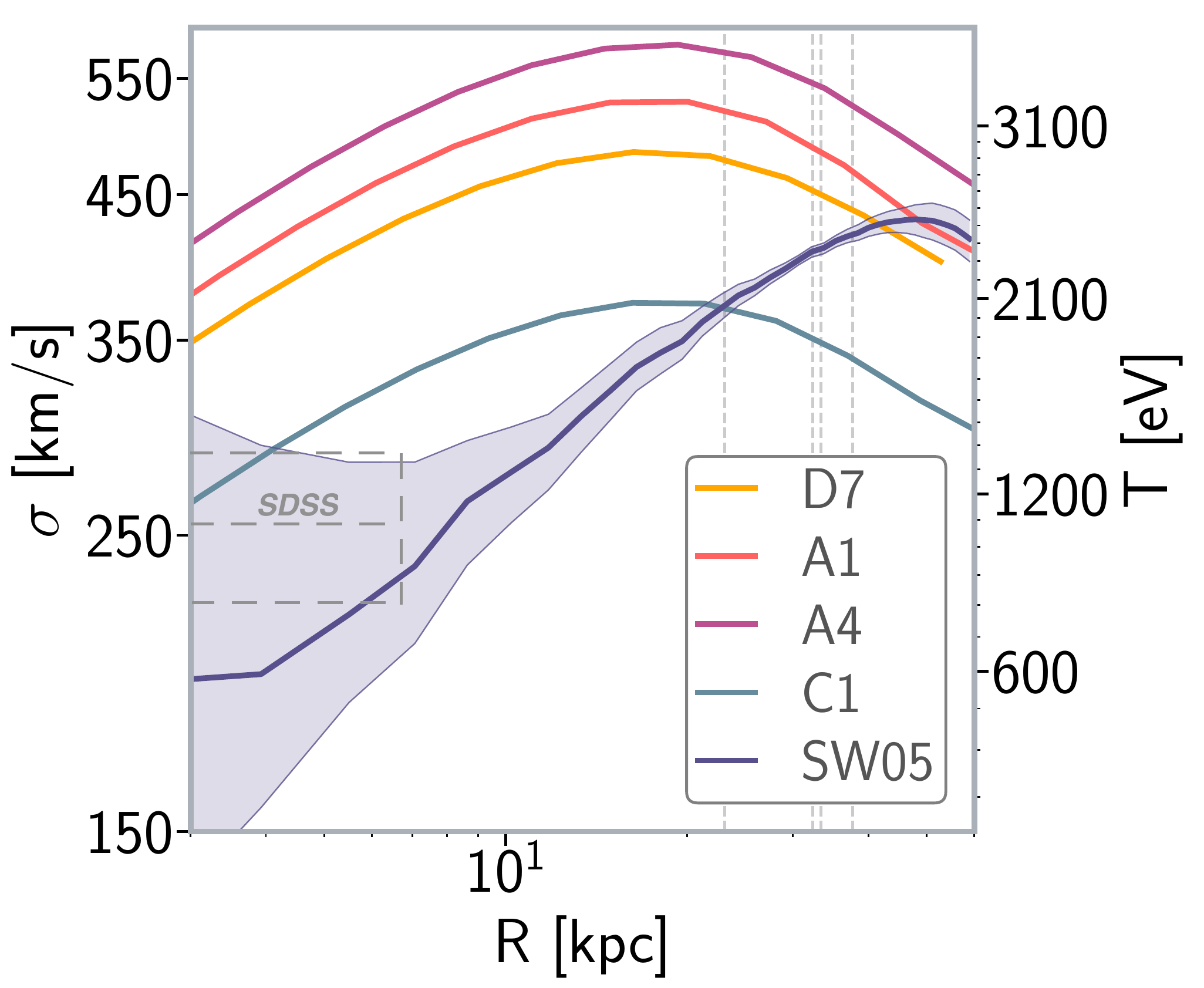}
  \caption[Derived velocity dispersion profiles of SW05 in comparison
  with FIRE galaxy models]{
    Velocity dispersion profile of SW05, contrasted with
    galaxies from simulations: $\sigma$ is formally derived from the
    mass profiles shown in \figref{fossil:profiles} according to
    \eqref{fossil:dispersion}.  On the right y-axis the corresponding
    temperature values are denoted in units of $\mathrm{eV}$ according
    to \eqref{fossil:gas_temp}. The grey dashed lines mark the observed
    estimate from the SDSS/BOSS spectrum ($\sigma$=256$\pm$32\,km/s)
    and extends horizontally to map the fibre of the spectrograph.
  }\figlbl{fossil:dispersion}
  \end{figure}

%% file: fig/abmatch.tex
\begin{figure}
\includegraphics[width=\pwidth]{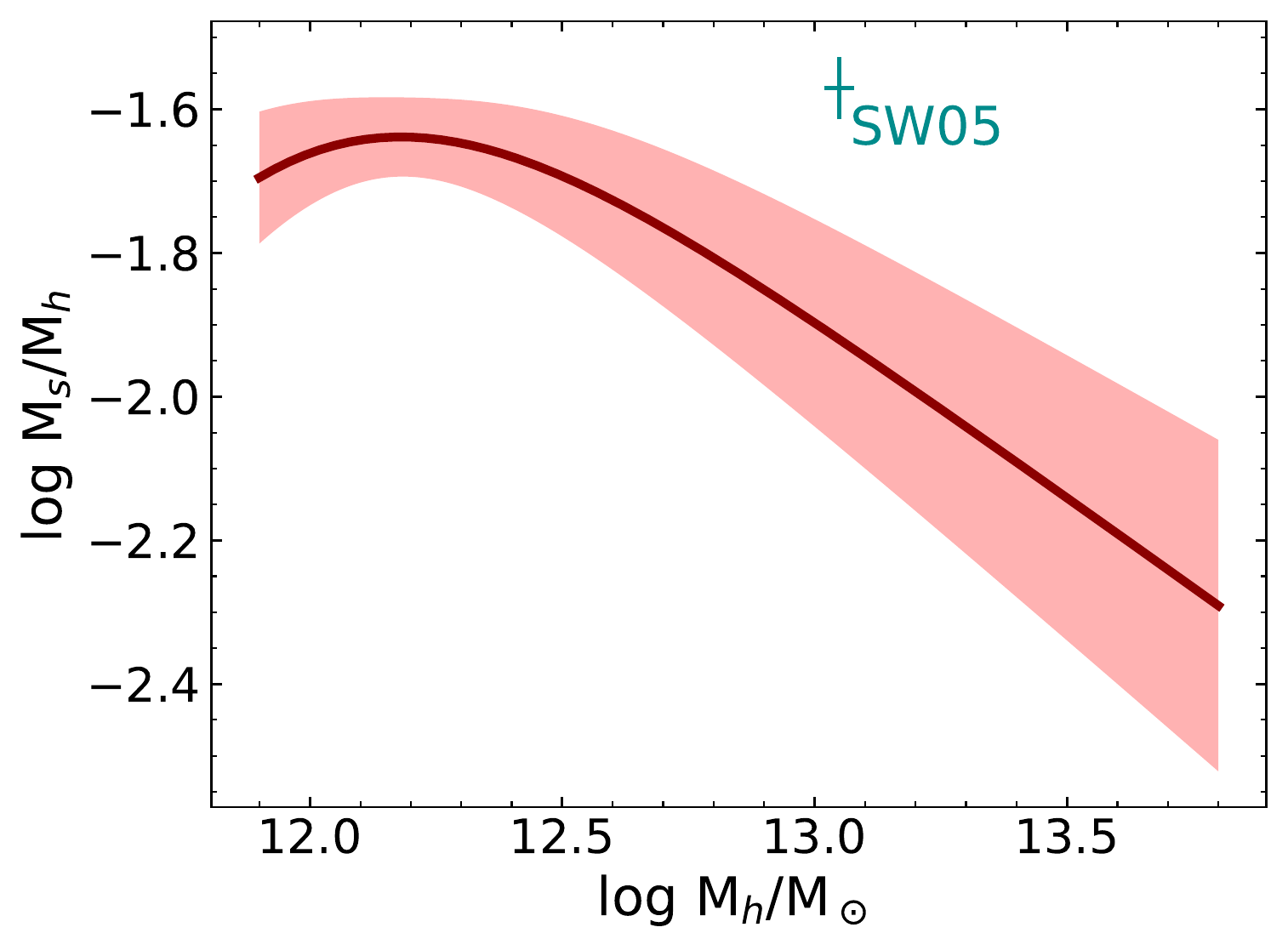}
\caption[SW05 in comparison with abundance matching]{
  The stellar to dark matter mass fraction is shown as a
  function of the halo mass. The solid line and (1\,$\sigma$) shaded
  region represent the standard abundance matching relation, at the
  redshift of the lens z$_L$=0.625, taken from \citet{Moster:10}. The
  error bars for the SW05 lens are given at the 1\,$\sigma$ confidence
  level.}\figlbl{fossil:AM}
\end{figure}

%% file: tex/appendix.tex
\appendix

\section{Early models}

As mentioned in \secref{fossil:system} the SW05 was discovered in the \SW
citizen-science project, appearing as one of 29 promising lens
candidates in \cite{SW2}.  As an especially interesting candidate,
SW05 had been discussed in the \SW online forum since 2013, under the
name ASW0007k4r, which was the \SW internal name for the particular
field containing the object.  \figref{fossil:7k4r} shows some early
results discussed on the forum.  As well as a false-colour image that
highlights the possible lensed images, there are models for caustics
indicating a near-fold system, an arrival-time surface, and histograms
of model time delays between consecutive images, in the event that the
source galaxy has a variable source.

\def\zwidth{.48\hsize}
\begin{figure}
\hbox to\hsize{%
\includegraphics[width=\zwidth]{\home 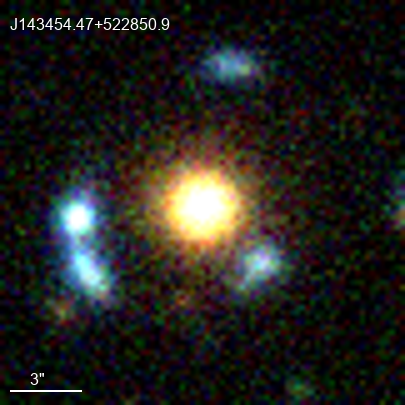}\hss
\includegraphics[width=\zwidth]{\home 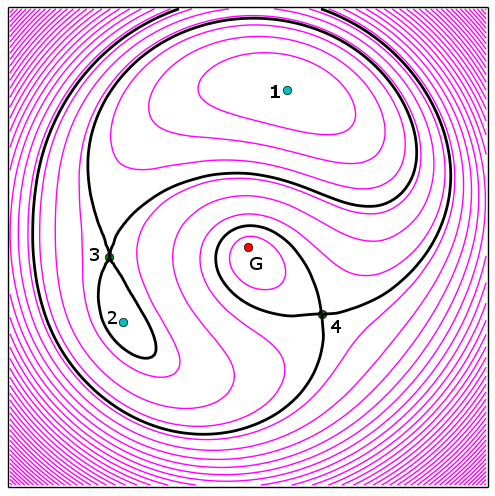}}
\hbox to\hsize{%
\includegraphics[width=\zwidth]{\home 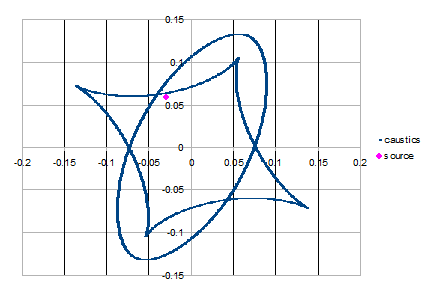}\hss
\includegraphics[width=\zwidth]{\home 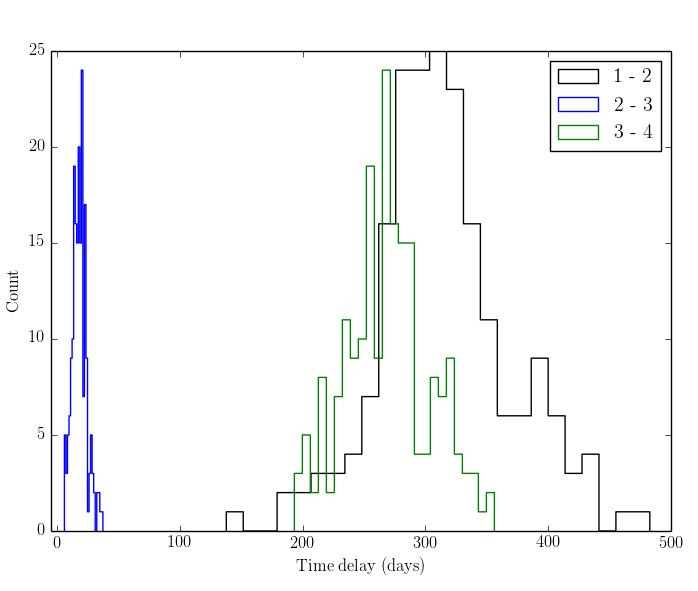}}%
\caption{Early results from \SW indicating that SW05 is indeed a
      lensing system.  Upper left:~a false-colour image enhancing the
      candidate lensed images.  Lower left:~Possible caustic
      configuration. Upper~right:~Arrival time surface, similar
      to \figref{fossil:arrival} indicating the inferred image
      order. Lower right:~Histograms of model time delays between the
      images. \figlbl{fossil:7k4r}}
\end{figure}